# An Intelligent Future Mobile Terminal Architecture

Muhammad Bilal
Department of Computer System Engineering,
N-W.F.P University of Engineering and Technology, Peshawar Pakistan
{engr.mbilal@yahoo.com}

*Abstract* — In this paper, a novel Extended Cognitive Mobile Terminal (ExCogNet-MT) scheme is presented. In this scheme, a "test bench" at receiver's Mobile Terminal (MT) can estimate the channel Signal to Noise Ratio (SNR) and can detect the jamming signal. The estimation scheme compares the Standard Deviation (SD) of received signal and processed signal, and on the bases of this comparison the "test bench" can determine the BER and corresponding SNR value of the channel. Simulation results demonstrated that under certain scenarios estimated SNR value can be helpful for tuning the parameters of protocol stack of 802.11a and WiMaxm.

## I. INTRODUCTION

The future mobile terminals will have the ability to adapt different available existing and future network standards. From mobile station perspective, there should be a Multimode Mobile Station capable of adapting different wireless networks preferably by reconfiguring itself i-e turning hardware problems into software problems. Therefore, an intelligent scheme at mobile terminal is required to handle the inter technology switching. Some of the inter technology switching scenarios are: (1) The subscriber starts using an application that can be run efficiently on another available network. This new application may require more bandwidth/data rate for best performance then the data rate of current selected network and mobile sense the high-speed network. Or this new application requires less bandwidth/data rate for best performance then the data rate of current selected network and mobile sense the high congestion or some error in the selected network. In both situations the mobile terminal will go through the process of inter technology switching. Or the mobile terminal selects the best suitable network for the running application and the subscriber starts moving from its position and reaches at certain point where mobile terminal detects new best suitable network. (2) The mobile terminal selects the best suitable network for the running application and MT detects that signal strength of current selected network is weak and unable to find the neighbor cell of selected network in that coverage area. The MT will scan for the $2^{nd}$ best available network if it fails it will try for $3^{rd}$, $4^{th}$ and so on. (3) When a MT running the application on the best selected network and the network detects that the billing account for the selected network is over. The network will inform the MT by sending "End-of-Balance" and will give certain amount of time to switch over to other network with running application. If MT continues after reception of "End-of-Balance" packet the network will automatically disconnect the MT. (4) In idle mode, when MT detects that the signal strength of the selected network is weak and unable to find the neighbor cell of selected network in that coverage area. The MT will scan for the network which gives largest converge area or which have largest cell size.

(5) In the cognitive radio networks, the users are divided into two groups; [5]

- Primary user group: These users have the high priority to access the licensed part of the spectrum that is available in the operating region. In other words, these users are customers of the mobile operators who hold that licensed part of the spectrum.

- Secondary user group: These users have the low priority to access the licensed part of the spectrum that is available in the operating region. In other words, these users are not customers of the mobile operators who hold that licensed part of the spectrum. The secondary users can opportunistically access the licensed portion of spectrum through wideband access technology provided that the primary users face zero or little interference.

If Cognitive user is using a portion of spectrum and at the same time the same part of spectrum is required by the primary user. The MT will scan the spectrum to find out the second best available network if it fails it will try for $3^{rd}$, $4^{th}$ and so on.

The inter technology switching scenarios given in point 1 to 3 and point 5 are also known as "Vertical Handover". While the inter technology switching scenario # 3 is network initiated and 1, 2 and 4 are MT initiated inter technology switching. Both scenario # 02 and #04 depend upon the signal strength of the selected network and decision for inter technology switching takes place at PHY-layer. Scenario # 01 is related to the network optimization and Quality of Service (QoS) of running application.

Whether the inter technology switching is MT initiated or Network initiated the most important thing is, this switching time should be minimized to avoid the handover latency delay. The latency delay for scenario # 02 and #04 can be minimized if the MT has in advance knowledge of link quality.

The channel signal to noise ratio (SNR) is the most important parameter for measuring the link quality. The estimated value of channel SNR can play a crucial role in self reconfiguration of MT. It is very important to find out a suitable SNR estimation scheme. A. Ramesh et al [2]t presented an SNR estimation technique for the generalized fading channel which requires no polite or training bits. Their simulation showed that this scheme can work very accurately for the turbo decoders in the Nakagami fading channel. The single point estimated SNR value cannot reflect the actual SNR, therefore pattern of estimated SNR for certain interval of time along with corresponding bandwidth can reflect the actual SNR value [3]. Moreover, if the MT has the estimated value of channel SNR, the applications and the protocols can effectively manage the resources. The next sections of this paper present a mobile terminal model which can effectively estimate the Bit Error Rate (BER) and corresponding channel SNR with in an acceptable range. The aim is not just to find out SNR estimation scheme but to use the knowledge of estimated BER and SNR value in relation with Standard Deviation (SD) to make vertical handover decision in time and to tune the parameters of other protocol layers.

The recent development in the field of cognitive networking and the software defined radios has opened many paths to cope with challenges to the future mobile networks. Many projects of cognitive network architectures have been proposed and developed independently. In comparison with some famous cognitive network architectures; the CogNet is the only proposed architecture which is consistent with the TCP/IP protocol stack [7]. These projects were focused on the development of Cognitive Network Architecture but none of them discussed the cognitive node elements separately. This paper will discuss in detail the Cognitive mobile terminal architecture. Next section presents the architecture and working of user mobile terminal using cognitive networking approach.

II. INTELLIGENT FUTURE MOBILE ARCHITECTURE

A proposed architecture of cognitive complete knowledge system-CogNet is discussed by B.S. Manoj et al [4]. In the proposed model the Cognitive Node utilizes its own experiences and information (based on the experiences of other Cognitive Node gathered from all other cognitive nodes) to tune the parameters of protocol stack to optimize the overall network performance and to facilitate the Cognitive Node to take important decisions. The cognitive modules are distributed at each layer and can communicate via cognitive cross layer bus. The cognitive plane collects the observed events of cognitive modules at each layer and stores this spatiotemporal information in local repository. This stored information can be shared with other cognitive nodes. The reasoning algorithm analyzes the spatiotemporal information and takes the decision. The Cognitive radio system is responsible for adaptation of physical layer in according to the environment conditions.

Fig-1 represents the extended CogNet model for mobile terminal (ExCogNet-MT). In this model some new blocks are introduced to make it possible to run in a practical environment.

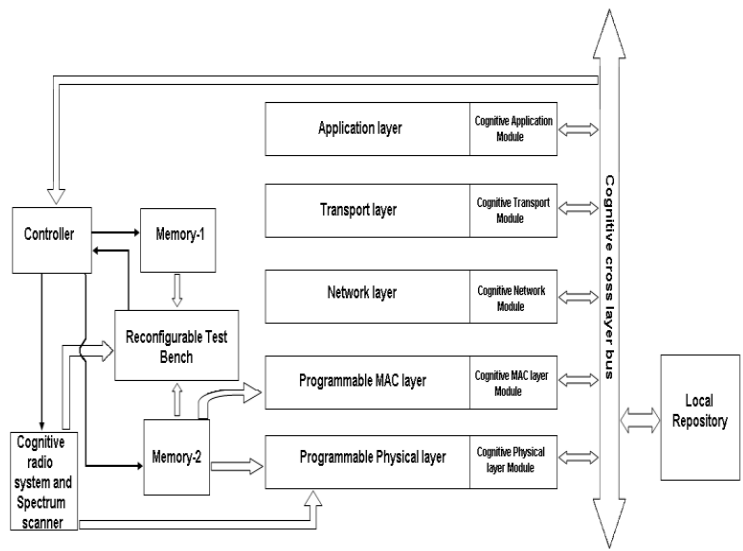

Figure 1. Extended CogNet Model

A. *Memory-1*

In this mobile terminal model, Memory-1 is a nonvolatile EEPROM. This memory is used to store the threshold values. These threshold values are predefined values, related to the behavior of different "wireless network standards" in different scenarios. For example, it can have the minimum acceptable BER and corresponding SNR value and the corresponding value of SD for different wireless network standards.

*B. Memory-2*

This is also a nonvolatile EEPROM, but this memory should be faster than memory-1 and should have larger data bus. This is used to store the HDL programs of physical layer and information of programmable MAC layer for different wireless network standards.

*C. Cognitive radio system and Spectrum scanner*

Whenever required the spectrum scanner, scans for the free spectrum holes and selects the portion of available spectrum such that the primary users should not be have an effect on. The ExCogNet-MT can share the information regarding spectrum condition via central repository and local repository. Instead of scanning complete spectrum (0~6 GHz), usually the spectrum scanner scans some part of the spectrum according to the running application.

*D. Controller*

The reasoning and decision-making algorithm are running in this part of the architecture. The controller receives the input from the test bench, from the cognitive module of all layers and can access the spatiotemporal data stored in the local repository. On the basis of all inputs and the spatiotemporal data of local repository, the controller instructs the layers of protocol stacks to tune the parameters for optimum utilization of resources and to maintain the QoS.

*E. Reconfigurable test bench*

The reconfigurable test bench is used to estimate the channel BER and inform the controller if the value of BER approaches to the maximum acceptable value. The test bench can adopt any mechanism to estimate the channel SNR e.g Time series modeling method of measuring the channel SNR given by J. Zhang et al [3].

Fig-2 represents the Reconfigurable test bench model which is used in the ExCogNet-MT model. This model is using the "Re-Modulation" scheme to estimate the BER and then corresponding channel SNR. Just like physical layer the Test Bench demodulates and decodes the input data coming from the Cognitive Radio System, but it also encodes and modulates the recovered data at receiver side i-e Re-Modulation. The functionality of different elements of "Reconfigurable Test Bench" is discussed below.

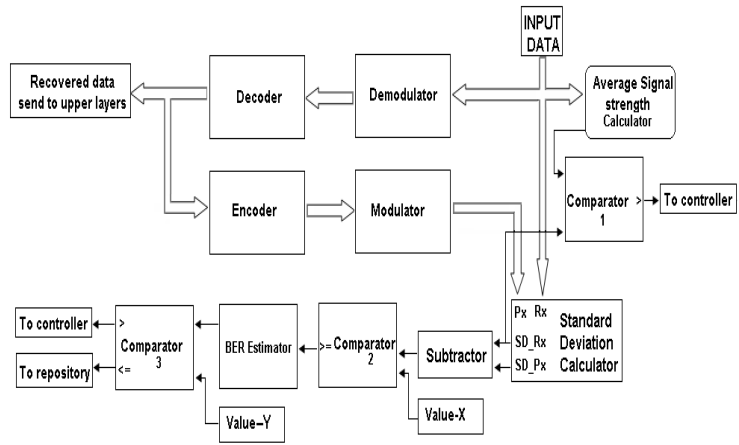

Figure 2.  Reconfigurable Test Bench

*1) Average Signal Strength*

By using the method of determining the difference average signals strength discussed by K. Reese et al [1], the "Average Signal Strength calculator" calculates the average signals strength and sends it to Comparator-1.

*2) Standard deviation calculator*

The "standard deviation calculator" calculates the standard deviation of input data and Re-Modulated data.

*3) Comparator-1*

The "Comparator-1" compares the value of standard deviation and the average signal strength of input data.

*4) Subtractor*

The "Subtractor" takes the absolute difference of the standard deviations of input data and Re-Modulated data.

*5) Value-X*

Value-X is a predefined threshold value which is stored in a register. This value indicates an acceptable threshold for the absolute difference of the standard deviations of input data and Re-Modulated data. This can be obtained by simulating the "Test Bench" for different wireless standards and choosing value-x on hit and trail basis. Appendix-A shows the statistical data of simulation results it is found that to estimate the BER/SNR in acceptable range "the threshold absolute difference of the standard deviations" for WLAN 802.11a is 0.00205 and for WiMAX it is 0.0048.

 6) *Comparator-2*

The "Comparator-2" compares the result of "Subtractor" with "Value-X".

 7) *BER Estimator*

The "BER Estimator" estimates the BER (bit error rate) on the basis of results obtained from "Comparator-2". Fig-3 shows the simulation results of test bench for WLAN 802.11a and Fig-4 shows the result of simulation of test bench for WiMAX. In both figures thesolid lines represents the actual values and dashed lines represents the estimated values.

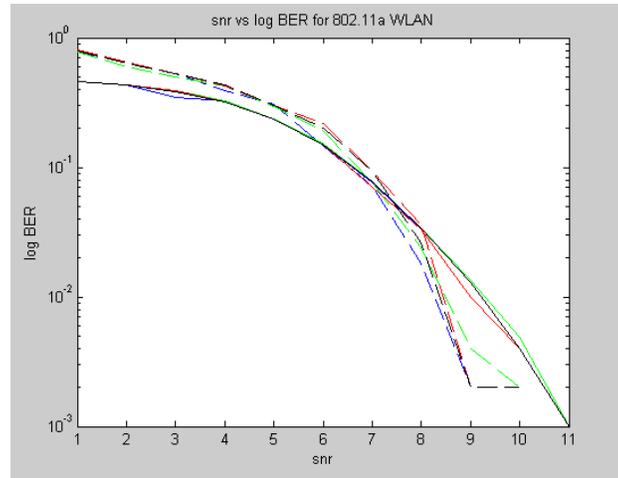

Figure 3.  WLAN SNR Vs $Log_{10}$ (BER)

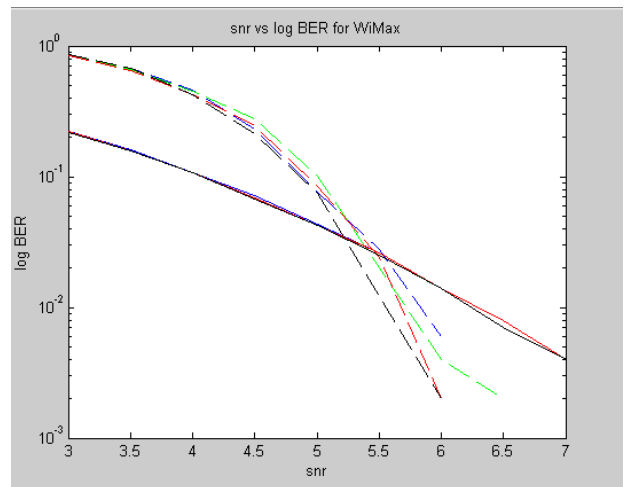

Figure 4.  WiMAX SNR Vs $Log_{10}$ (BER)

From the above figures it is clear that the estimated BER is rapidly increasing from actual values at SNR below "5". The SNR 5 can be set as an acceptable minimum threshold for WiMAX and SNR 7 can be set as an acceptable minimum threshold for WLAN 802.11a.

 8) *Value-Y*

Value-Y is a predefined threshold value which is stored in a register. This value indicates an acceptable threshold for the acceptable BER.

 9) *Comparator-3*

The "Comparator-3" compares the result of "BER Estimator" with "Value-Y".

### III. WORKING AND OPERATION:

The following algorithm explains the usage of Test Bench and operation of "Extended CogNet Mobile Model".

```
Variables:
D =  difference average signal strength [1].
SD_Rx  = standard deviation of input received data.
SD_Diff = Result of Subtractor.
Value-X = This value indicates an acceptable threshold for the absolute difference of
the standard deviations of input data and Re-Modulated data.
Value-Y = This value indicates an acceptable threshold for the acceptable BER and SNR.
Est_BER = Out put of BER Estimator.
Erroneous_Frame  =  Number of frames detected in error.
Total_Frams = Total Number of frames Re-Modulated.

Algorithm:
   1. REPEAT-1: line 2 to 12 forever
   2. IF   D>SD_Rx
   3.          Inform Controller
   4. ELSE
   5. REPEAT-2: line 5 to 6 for 500 times
   6. IF  SD_Diff >= Value-X
   7.          Erroneous_Frame  = Erroneous_Frame  +1
   8. END REPEAT-2
   9. Est_BER =  Erroneous_Frame  / Total_Frams
   10. IF  Est_BER<= Value-Y
   11. Send Estimated value to local repository.
   12. ELSE IF  Est_ BER>Value-Y
   13.           Inform Controller
   14. END REPEAT-1
```

In line-2 if the "Comparator-1" finds that the value of average signal strength of input data is greater than standard deviation, it informs the controller; in response the controller instructs the cognitive radio system to broad cast the "jamming signal detected" message.

Line-4 is start of loop which takes 0.05sec and will repeat line 5 to 6 to check 500 frames. In line-5 if the "Comparator-2" finds that the result of Subtractor is greater than value-x, it means this frame contains unrecoverable errors, so it will increase value of variable Erroneous_Frame by "1". In line-8 we get the estimated value of BER by dividing the number of erroneous frames detected in step-4 to 7 and total number of frames, which is 500 in this case. From line-9 to 12 the Comparator-3 compares estimated value of BER with value-y. If BER is less than or equal to Value-Y the comparator informs the controller and the controller will send the estimated value to local repository via cognitive cross layer bus. In repository BER will store along with corresponding SNR value by using "MAP function" where BER (rounded off to certain digits) will work as an input for MAP function. The MAP function is a kind of operation in which it maps the BER to the corresponding SNR by finding the closest BER in lookup table. The lookup table for different standards can be built by simulation results. The tables in Appendix-A for WLAN and WiMAX both can be used as a lookup table for WLAN and WiMAX. If estimated BER is greater than Value-Y it means that channel conditions are going worst and it's the time to handoff horizontally or vertically. The controller will instruct the "Cognitive radio system and Spectrum scanner" to search for the available spectrum holes. The controller will select the most suitable spectrum portion. This selection depends upon the QoS required for running application and the right of customer to use licensed portion of spectrum. Quality of service varies from application to application; required QoS of various applications is summarized by P. Newton et al [6]. Once the decision for selection of suitable spectrum has taken the Controller will instruct the Memeory-2 to reconfigure the physical layer, MAC layer and Test Bench according to selected spectrum portion. Once reconfiguration is done the controller will instruct the Memory-1 to transfer appropriate required parameter to test bench e.g Value-X and Value-Y.

In section-I the four inter technological switching scenarios are discussed. The scenario # 2 and 4 are dependent upon the signal strength of the selected network. Our mobile architecture not only detects the minimum acceptable BER and corresponding SNR of channel but also keep track of BER/SNR which helps the MT to make inter technological switching decision before the reaching the worst condition. The track record of BER/SNR stored in local repository can be shared among the cognitive modules of all layers and with other cognitive nodes, especially when a new MT enters in the network. The scenario # 1 is related to the QoS of running application. The Application layer can inform the Controller about the QoS requirements and the current situation of network, in response the Controller will instruct the cognitive radio system to find the best suitable network/spectrum portion.

## IV. CONCLUSION

This paper discussed the architecture of future mobile terminal and explained its functionality for few scenarios. The basic concept of this architecture is to utilize the cognitive spatiotemporal information in such a way that the challenges faced by the future mobile network technologies due to the network heterogeneity can be solved. In future the functionality of test bench should be enhanced to estimate the channel condition for different fading processes other then AWGN and multipath interference phenomena should also be included. In this Architecture test bench use to process '500' frames and in each 1 sec total time for running the test bench algorithm is 0.05 sec. The number of frames to process and time for running the test bench algorithm depends upon the type of network. The appropriate number of frames to process and total time for running the test bench algorithm should be selected for different networks after considering different statistical and probabilistic data gathered for different situations. Similarly, different threshold, test bench values e.g Value-X & Y for different networks should be determined by simulation results.

APPENDIX-I

Table 1 & 2 shows the statistical data, obtained from the simulation results of testbench. Both tables can be used as a lookup tables for value-x and value-y (discussed in section-II) and also the algorithm (discussed in section-III) is using the table values. Data has been collected for different channel conditions i-e the different SNR values of AWGN channel and for different Data patterns. The diverse Data patterns are generated by a random Bernoulli binary generator by using different seed values.

Estimated BER= Number of erroneous frames detected /total frames

**WLAN: 802.11a**

Data rate is ~=21Mbps

Total number of frames 500 and simulation time 0.05sec, threshold Standard Deviation difference is 0.00205.

**Table 1 Statistical Data of WLAN802.11a**

| SNR AWGN Channel | SEED values of data source | | | | | | | | | | | |
|---|---|---|---|---|---|---|---|---|---|---|---|---|
| | 10 | | | 40 | | | 70 | | | 100 | | |
| | BER | | Number of erroneous frames detected | BER | | Number of erroneous frames detected | BER | | Number of erroneous frames detected | BER | | Number of erroneous frames detected |
| | Actual | Estimated | | Actual | Estimated | | Actual | Estimated | | Actual | Estimated | |
| 1 | 0.46 | 0.774 | 387 | 0.457 | 0.77 | 385 | 0.458 | 0.804 | 402 | 0.457 | 0.796 | 398 |
| 2 | 0.43 | 0.654 | 327 | 0.431 | 0.60 | 300 | 0.43 | 0.652 | 326 | 0.431 | 0.63 | 315 |
| 3 | 0.349 | 0.528 | 264 | 0.389 | 0.498 | 249 | 0.388 | 0.53 | 265 | 0.387 | 0.524 | 262 |
| 4 | 0.323 | 0.392 | 196 | 0.324 | 0.424 | 212 | 0.322 | 0.424 | 212 | 0.321 | 0.432 | 216 |
| 5 | 0.238 | 0.308 | 154 | 0.238 | 0.294 | 147 | 0.238 | 0.298 | 149 | 0.238 | 0.302 | 151 |
| 6 | 0.148 | 0.144 | 72 | 0.15 | 0.19 | 95 | 0.147 | 0.218 | 109 | 0.148 | 0.202 | 101 |
| 7 | 0.076 | 0.072 | 36 | 0.078 | 0.076 | 38 | 0.07 | 0.094 | 47 | 0.078 | 0.094 | 47 |
| 8 | 0.033 | 0.018 | 9 | 0.034 | 0.024 | 12 | 0.033 | 0.036 | 18 | 0.034 | 0.026 | 13 |
| 9 | 0.01 | 0.002 | 1 | 0.0134 | 0.004 | 2 | 0.01 | 0.002 | 1 | 0.013 | 0.002 | 2 |
| 10 | 0.004 | 0 | 0 | 0.0049 | 0.002 | 1 | 0.004 | 0.002 | 1 | 0.004 | 0.002 | 1 |
| 11 | 0.001 | 0 | 0 | 0.001 | 0 | 0 | 0.001 | 0 | 0 | 0.001 | 0 | 0 |

**WiMAX:**

Data rate is ~=3Mbps

Total number of frames 500 and simulation time 0.05sec, threshold Standard Deviation difference is 0.0048.

**Table 2 Statistical Data of WiMax**

| SNR AWGN Channel | SEED values of data source | | | | | | | | | | | |
|---|---|---|---|---|---|---|---|---|---|---|---|---|
| | 10 | | | 40 | | | 70 | | | 100 | | |
| | BER | | Number of erroneous frames detected | BER | | Number of erroneous frames detected | BER | | Number of erroneous frames detected | BER | | Number of erroneous frames detected |
| | Actual | Estimated | | Actual | Estimated | | Actual | Estimated | | Actual | Estimated | |
| 3 | 0.22 | 0.85 | 420 | 0.221 | 0.848 | 424 | 0.22 | 0.85 | 420 | 0.217 | 0.852 | 426 |
| 3.5 | 0.16 | 0.67 | 338 | 0.158 | 0.664 | 332 | 0.159 | 0.65 | 325 | 0.157 | 0.67 | 335 |
| 4 | 0.108 | 0.456 | 228 | 0.107 | 0.452 | 226 | 0.107 | 0.428 | 214 | 0.108 | 0.422 | 211 |
| 4.5 | 0.071 | 0.232 | 116 | 0.069 | 0.276 | 138 | 0.069 | 0.244 | 122 | 0.068 | 0.214 | 107 |
| 5 | 0.043 | 0.076 | 38 | 0.042 | 0.102 | 51 | 0.042 | 0.084 | 42 | 0.042 | 0.074 | 37 |
| 5.5 | 0.026 | 0.028 | 14 | 0.025 | 0.02 | 10 | 0.026 | 0.024 | 12 | 0.025 | 0.012 | 6 |
| 6 | 0.014 | 0.006 | 3 | 0.014 | 0.004 | 2 | 0.014 | 0.002 | 1 | 0.014 | 0.002 | 1 |
| 6.5 | 0.008 | 0 | 0 | 0.007 | 0.002 | 1 | 0.008 | 0 | 0 | 0.007 | 0 | 0 |
| 7 | 0.004 | 0 | 0 | 0.004 | 0 | 0 | 0.004 | 0 | 0 | 0.004 | 0 | 0 |